\documentclass[apj]{emulateapj}

\def\CaI{\hbox{Ca \sc i}}      
\def\CaII{\hbox{Ca \sc ii}}
\def\OI{\hbox{O \sc i} }

\def\HeI{\hbox{He \sc i}}
\def\FeII{\hbox{Fe \sc ii}}
\def\CrII{\hbox{Cr \sc ii}}
\def\TiII{\hbox{Ti \sc ii}}
\def\FeIII{\hbox{Fe \sc iii}}
\def\AlIII{\hbox{Al \sc iii}}
\def\CII{\hbox{C \sc ii}}

\def\NaI{\hbox{Na \sc i}}
\def\kms{km s$^{-1}$}
\def\ang{$\hbox{\AA \,}$}
\def\ha{H$\alpha$}
\def\hb{H$\beta$} 
\def\coude{Coud\'{e}}
\def\micron{$\mu$m}

\begin{document}
\title{Lifting the Iron Curtain : \\ Toward an Understanding of the Iron Stars XX Oph and AS 325}
\author{Richard J. Cool\altaffilmark{1}, Steve B. Howell\altaffilmark{2,3}, Maria Pe\~{n}a\altaffilmark{3,1}, Andy J. Adamson\altaffilmark{4}, and Robert R. Thompson \altaffilmark{5}}

\altaffiltext{1}{Steward Observatory, University of Arizona, 933 N. Cherry Ave., Tucson, AZ 85721; {\tt rcool@as.arizona.edu}}
\altaffiltext{2}{WIYN Observatory, 950 North Cherry Avenue, Tucson, AZ 85726; {\tt howell@noao.edu}}
\altaffiltext{3}{National Optical Astronomy Observatory, 950 North Cherry Avenue, Tucson, AZ 85726; {\tt  mapeague@email.arizona.edu}}
\altaffiltext{4}{UK Infrared Telescope Joint Astronomy Centre, 660 N A'Ohoku Place, Hilo, HI 96720, USA; {\tt a.adamson@jach.hawaii.edu}}
\altaffiltext{5}{Michelson Science Center, MS 100-22, California Institute of Technology, 770 South Wilson Drive, Pasadena, CA 91125; {\tt thompson@ipac.caltech.edu}}

\begin{abstract}

We present new optical, near infrared, and archival ultraviolet observations of XX Ophiuchi and AS 325, two proposed ``iron'' stars.  These unusual stars have optical spectra dominated by emission lines arising from hydrogen as well as ionized metals such as iron, chromium, and titanium.  Both stars have been classified as ``iron'' stars and a number of exotic models have been presented for their origin.  Using two years of moderately high resolution optical spectroscopy, the first high signal-to-noise K-band spectroscopy of these sources (which reveals stellar photospheric absorption lines), and new near-infrared interferometric observations,  we confirm that both systems are composed of two stars, likely binaries, containing a hot Be star with an evolved late-type secondary.  The hydrogen emission features arise in the hot wind from the  Be star while the corresponding P-Cygni absorption lines are produced from dense material in the expanding, radiation driven, wind around each system.  The optical \FeII\, emission lines are pumped by ultraviolet \FeII\, absorption lines through fluorescence.   Contrary to some claims in the literature, the spectral features of XX Oph and AS 325 are quite similar, evidence that they are comparable systems. We examine the  variability of the spectral morphology and radial velocity motions of both sources.  We also study the variability of XX Oph during a major photometric event and find that the spectral nature of the system varies during the event. A comparison of the velocity of the absorption line components in our new spectra with those in the literature  show that the structure of the stellar wind from XX Oph has changed since the system was observed in 1951.

\end{abstract}


\section{Introduction}

XX Ophiuchi  was noted to be peculiar with its classification as an ``iron'' star at the beginning of the twentieth century \citep{m1924}. Using photographic spectra, \citet{merrill1951} found 577 emission features between 3600 and 6500 \ang, only 10 of which he did not identify. The lines consist primarily of hydrogen and ionized metals such as \FeII, \CrII, and \TiII \, with some contribution from  \hbox{V \sc ii}, \hbox{Sc \sc ii}, \hbox{Ni \sc ii}.  While the peculiar nature of this star was emphasized, no explanation was offered for the origin of the emission in the system. 

\citet{grl2001}  obtained high resolution spectra of XX Oph over a year long baseline and found that the emission lines exhibit peculiar variabilities compared to the original observations used to classify the system as an ``iron'' star  in 1924.  While no firm conclusion about the physical nature of the star was drawn, the authors suggest  the structure and mass loss in XX Oph may be explained through a binary interaction.  \citet{dt1990}  argue the presence of Balmer and \HeI \, absorption indicates the system contains a B0-B1e primary and a near-infrared excess arises due to the presence of an M6 III secondary. These claims are backed by the presence of strong \FeII\, absorption against an underlying continuum in the UV \citep{mich1991} and the presence of TiO bands in the near infrared \citep{hg1977}.  \citet{evans1994} discovered signatures of polycyclic aromatic hydrocarbons (PAHs) in the stellar atmosphere which suggest an inferred cool component to XX Oph is either carbon rich or oxygen rich with an external source of PAH formation. XX Oph is a photometric variable, as well. The star has historically had intermittent, one magnitude, drops in its optical flux lasting several years \citep{prager}.  In March 2004, the flux of the star decreased by 1.5 magnitudes, reaching its faintest state in 37 years \citep{sobotka2004}. No clear correlation between the spectral and photometric variability has been found; XX Oph experienced a  photometric minimum between 1921 and 1922 with no accompanying spectral variations \citep{m1924}.

The spectral similarities of the stars AS 325  and  XX Oph were first noted by \citet{bh1989}.  The spectra of both stars show strong hydrogen emission lines as well as emission from ionized metals; neither star has a normal stellar continuum in the optical. While XX Oph has a diverse history in the literature, AS 325 has received much less attention.  It was assigned a spectral type of A7 Ia+ pec in the \citet{sw1972} spectral survey and type F based on strong \CaII\,  absorption by \citet{ss1973}. \citet{M2002} tentatively classified AS 325 as a symbiotic star or related object.  The similarly of these two stars spawned the possibility that they are two examples of one type of object.  \citet{pereira2003} claim to see a  rising blue continuum in AS 325 and argue that this distinct difference from XX Oph indicates the two stars are not similar.

No past study has successfully attempted to consistently interpret all of the observed properties of these two systems across a range of wavelengths.  Also, the few infrared observations of these stars have not been used to constrain any possible models for their nature.  In this paper, we present new high resolution optical spectra, high signal-to-noise near infrared spectra, and  archival International Ultraviolet Explorer (IUE) spectra of XX Oph and AS 325 and use these observations to create one cohesive picture of these peculiar systems. In $\S$ 2, we present the data used in our analysis.  We discuss our findings in \S3 before concluding in $\S$4. 


\begin{figure*}[ht]
\epsscale{1.0}
\plotone{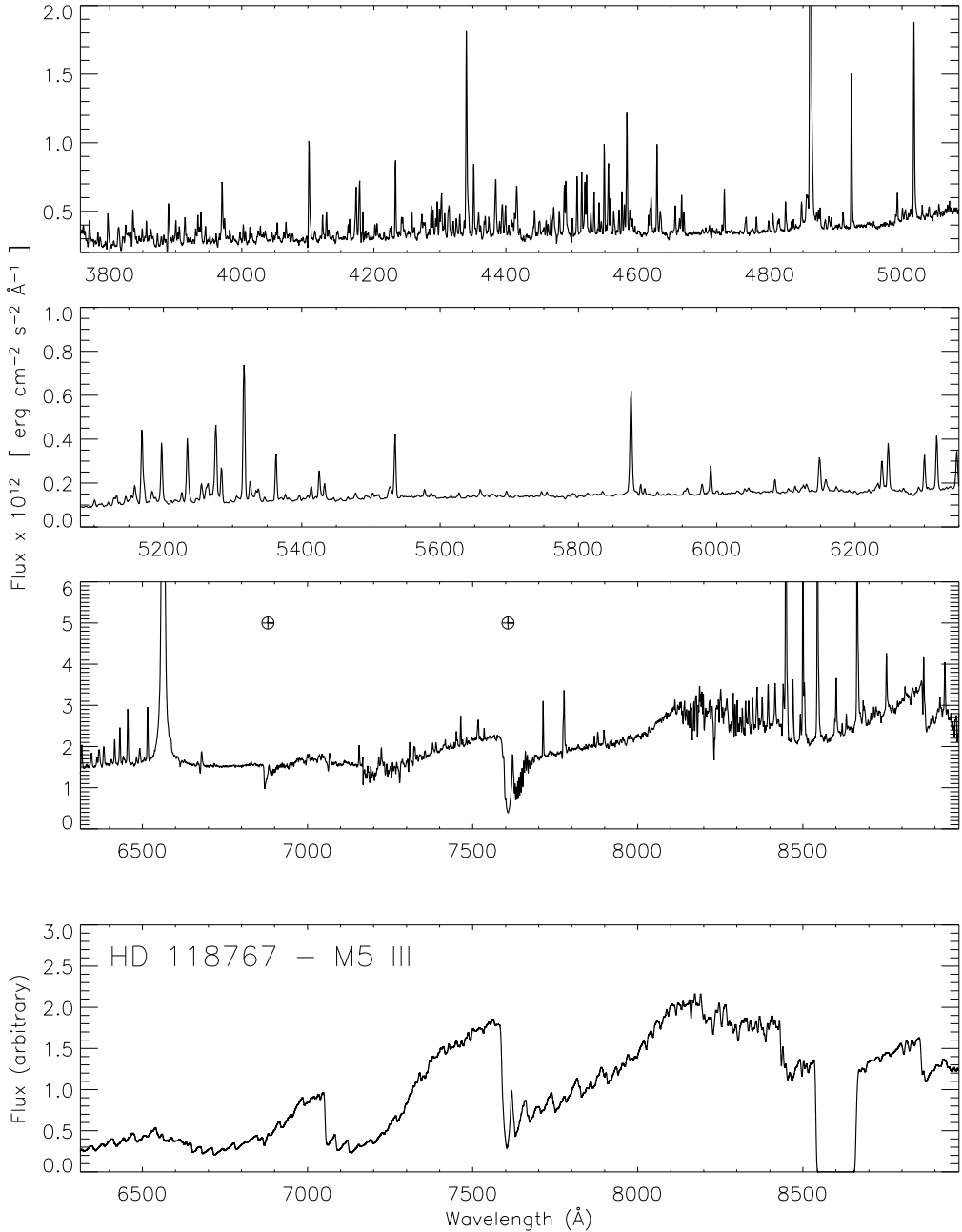}
\caption[]{ \scriptsize Optical spectrum of the star XX Ophiuchi and a comparison star of similar spectral type.  The strong absorption bands located in the red spectrum (and marked with the $\earth$ symbol) are telluric bands in the Earth's atmosphere and not inherent to the source. It should be noted that the spectra were taken at different epochs.  While the red and blue spectra were taken on 2003 July 05 and 2003 July 04, respectively, the green spectrum was taken on 2004 June  25.  The bottom panel shows an M5 III star, similar to our infrared determined M5 II classification, from the UVES High-Resolution library of stars which has been convolved to our resolution. The complex emission structure in the optical makes a direct comparison of the two spectra difficult, but there is a correspondence between the red continuum shape of the comparison star and that of XX Oph.  The sharp feature near 8600 \ang\, in the comparison spectrum is a region of missing data, not a true feature.}
\label{fig:xxophspec}
\end{figure*}

\section{Data}
Table 1 lists all of the observations used in the analysis presented here.  The details of the observations are summarized in the following sections according to the telescope used for the measurements.

\subsection{Kitt Peak Coud\'{e} Feed Telescope}

Spectra of each of the ``iron'' stars, as well as a sample of neighboring stars,  were obtained with the Kitt Peak National Observatory Coud\'{e} Feed Spectrograph between 10 June 2003 and 28 June 2004.  Each of the ``iron'' stars were observed in three configurations; a ``blue'' setup was used to observe the  $\lambda\lambda$ 3800-5100 \ang\,  range, an intermediate ``green'' setting was used to study the spectrum between \ha\, and \hb\, ($\lambda\lambda$ 4650-7000 \ang), and a ``red'' configuration, including \ha\,  and the Paschen limit, covered the spectral range $\lambda\lambda$ 6300-9000 \ang. The spectral resolution for each of these configurations  was 1-2 \ang and the spectra were generally obtained as back-to-back pairs of 300 second exposures which were combined to reject cosmic rays.  A sample of stars near both XX Oph and AS 325 was  observed with the blue configuration in order to search for spectral features similar to the two main stars of interest.   Data reduction, including bias and overscan subtraction, flat-fielding, wavelength calibration, and flux calibration was completed using standard packages in the Image Reduction and Analysis Facility (IRAF).

\subsection{Bok 2.3m Telescope}

Spectra of stars located spatially near XX Oph and AS 325, a deep spectrum of the blank sky near XX Oph, and a spectrum of AS 325 were obtained with the Boller and Chivens spectrograph on the Bok 2.3m Telescope operated by Steward Observatory.  The stellar spectra span $\lambda\lambda$ 3500-4500 \ang with a spectral resolution of 2 \ang while the blank sky spectrum covered 3350-5500 \ang with 3 \ang resolution.  Bias subtraction, flat-fielding, wavelength calibration, and flux calibration were completed using the {\tt iSPEC} package developed for IDL by John Moustakas.  {\tt iSPEC} not only provides a robust calibration and extraction routine for long slit spectra, but  also maintains the error on each measurement throughout the reduction process, resulting in a fully calibrated spectrum with meaningful errors on each pixel.

\newpage

\subsection{United Kingdom Infrared Telescope}
Near infrared spectra of XX Oph and AS 325 were obtained with the UIST spectrograph on the United Kingdom Infrared Telescope (UKIRT) on 2003 Oct 10. UIST is a 1-5\micron\, imager-spectrometer with a 1024x1024 InSb array and a number of grisms and slits available. Our single spectra used the HK grism and a long slit  0.48" wide at a plate scale of 0.12" pixel$^{-1}$. This  combination yielded a spectral resolution of R=800-1000 ($\sim$330 \kms)
across the 1.4 to 2.5 \micron\, region. The spectra were flux calibrated and telluric corrected in the normal way using spectral standard stars observed near in time and airmass to each of the program objects. At this spectral resolution, we cannot obtain meaningful velocity information but the spectra were key to our detection of the  stellar photosphere of the cool star in each system.

\begin{figure*}[ht]
\plotone{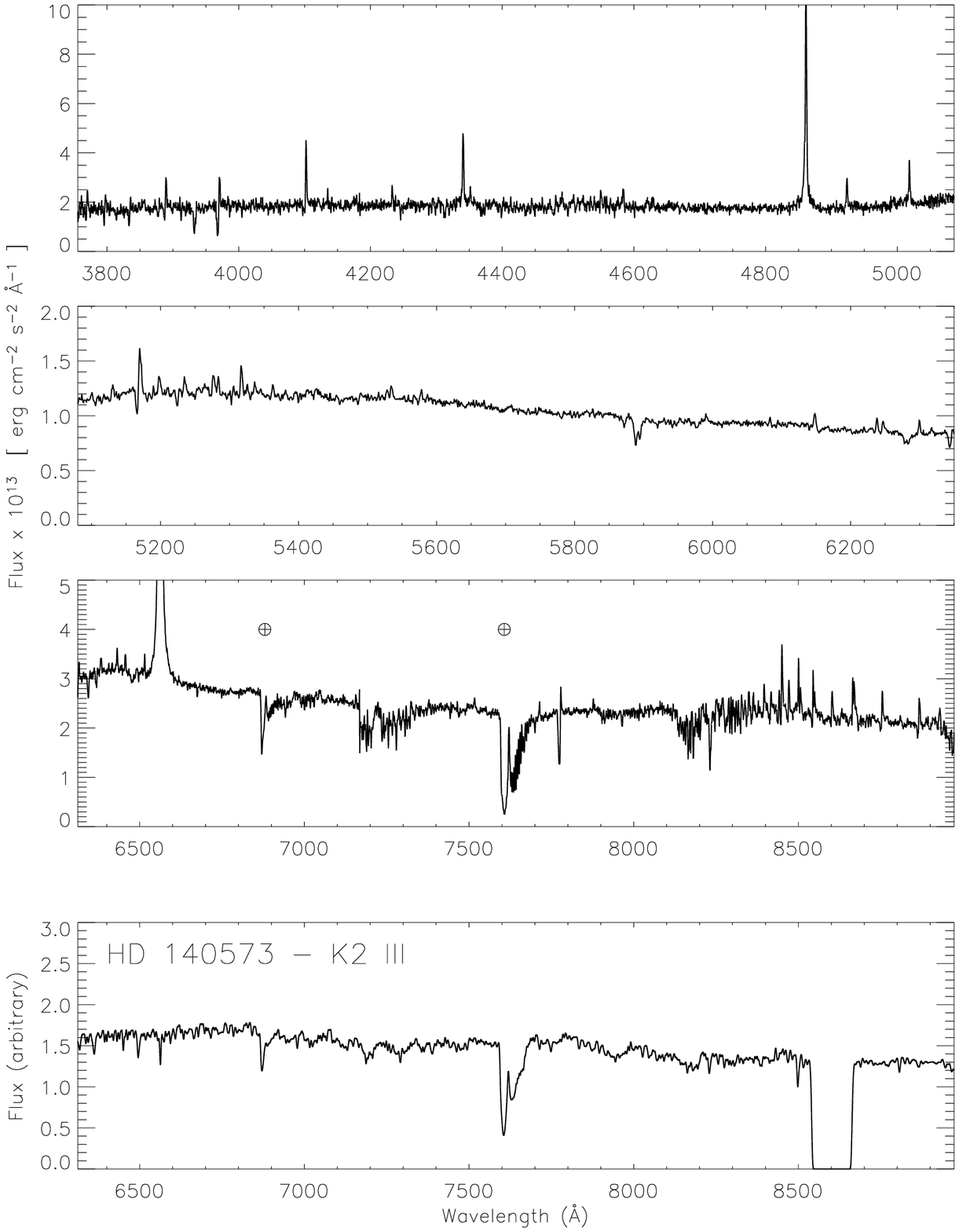}
\caption{ \scriptsize Same as Figure \ref{fig:xxophspec} but for the star AS 325 and a comparison K2 III star.}
\label{fig:as325spec}
\end{figure*}

\subsection{International Ultraviolet Explorer}
Ultraviolet spectra of XX Oph (SWP23783), AS 325 (SWP38427), and a comparison star, HD 87643 (SWP26040),  were taken from the IUE archive through the Multimission Archive at Space Telescope (MAST).  Each of these spectra were obtained in low dispersion mode ($R\sim270$) with the SWP camera and span the range $\lambda\lambda$ 1150-1980 \ang.  The stars were observed with a $10''\times 20''$ aperture.   Multiple observations are available in the archive for each of the stars; we present the spectra with the longest exposure times.  The spectra of XX Oph and HD 87643 have been previously published, thought for different purposes \citep{mich1991, pacheco}; the spectrum of AS 325 is previously unpublished.

\begin{deluxetable*}{lcclc}[
\tablecolumns{5}
\tablenum{1}
\tablewidth{0pt}
\tabletypesize{\scriptsize}                   
\tablecaption{Table of Observations}
\tablehead{
  \colhead{Object} &
  \colhead{Number of Spectra} & 
  \colhead{Date} & 
  \colhead{Observatory} & 
  \colhead{Spectral Range} \\
\colhead{} &
\colhead{} & 
\colhead{} &
\colhead{} & 
\colhead{(\AA)}
  }
\startdata
XX Oph & 1 & 26 August 1984 & IUE (SWP23783) & 1150-1980 \\
HD 876423 & 1 & 23 May 1985 & IUE (SWP26040) & 1150-1980 \\
AS 325 & 1 & 24 March 1990 & IUE (SWP38427) & 1150-1980 \\
XX Oph & 2&10 June 2003 & \coude & 6300-9000 \\ 
AS 325 & 2&10 June 2003 & \coude & 6300-9000 \\
XX Oph & 2&4 July 2003 & \coude & 3800-5100 \\
AS 325 & 3&4 July 2003 & \coude & 3800-5100 \\
XX Oph & 2&5 July 2003 & \coude & 6300-9000 \\
AS 325 & 2&5 July 2003 & \coude & 6300-9000 \\
XX Oph & 1&7 July 2003 & \coude & 6300-9000 \\
AS 325 & 2&7 July 2003 & \coude & 6300-9000 \\
XX Oph & 13&8 July 2003 & \coude & 6300-9000 \\
AS 325 & 13&8 July 2003 & \coude & 6300-9000 \\
XX Oph & 1&10 Oct 2003 & UKIRT & 15000-25000 \\
AS 325 & 1&10 Oct 2003 & UKIRT & 15000-25000 \\
AS 325 & 1&27 May 2004 & Bok & 3500-4500 \\
Nearby Stars$^*$ & 8&27 May 2004 & Bok & 3500-4000 \\
Blank Sky & 1&28 May 2004 & Bok & 3600-5000 \\
Nearby Stars$^*$ & 6&27 May 2004 & \coude & 3800-5100 \\
XX Oph & 1&25 June 2004 & \coude & 5200-6600 \\
AS 325 & 1&25 June 2004 & \coude & 5200-6600 \\
AS 325 & 2&26 June 2004 & \coude & 6300-9000 \\
AS 325 & 1&28 June 2004 & \coude & 3800-5100 

\enddata
\tablenotetext{*}{\scriptsize Table 4 lists the neighboring stars observed during these nights individually.}
\end{deluxetable*}

\section{Results and Discussion}

\subsection{Optical Emission}

Representative optical spectra of XX Oph and AS 325 are shown in Figures \ref{fig:xxophspec} and \ref{fig:as325spec} respectively.  The spectrum of each star is dominated by  hydrogen emission from the Balmer and Paschen series as well as ionized metal emission lines.  In AS 325, the metal emission lines are weaker than those in XX Oph.  These emission lines are accompanied by Balmer absorption lines blueshifted from the emission lines in classical P-Cygni profiles.  XX Oph shows P-Cygni profiles in \HeI \, while AS 325 contains absorption due to \CaII \, H+K and \NaI \, with no accompanying emission.  P-Cygni profiles are visible in all of the Balmer lines except H$\alpha$ which shows no clear absorption component.   The line profile of H$\alpha$ is quite asymmetric, in agreement with past studies in the literature \citep{merrill1951, grl2001}; the blue wing of the absorption line is truncated at the position of the expected absorption component. The lack of a clear absorption component probably indicates the \ha\, line is optically thick, allowing repeated scattering to obscure the absorption component, while the higher energy Balmer lines are optically thin. Table 2 lists identifications for some of the prominent lines in each of the stars based on the same spectra shown in Figures \ref{fig:xxophspec} and \ref{fig:as325spec}; the observed wavelengths listed in the table are based on a single observational epoch. \citet{pereira2003} suggest that AS 325 possesses a strong blue continuum. We find no strong blue continuum in any of our AS 325 spectra including spectra taken with different instrumental configurations which were reduced independently using different reduction software. The similarities between XX Oph and AS 325 are striking and a strong indication that the radiative mechanisms in these two stars share common physical processes.

It should be noted that the spectral fluxes between each panel in Figures \ref{fig:xxophspec} and \ref{fig:as325spec} show some variations.  In both stars, the ``green'' spectra, where were taken nearly a year later than the other spectra presented, appear offset from the ``red'' and ``blue'' observations.   In XX Oph, this difference agrees well with the photometric variability of the star (the ``green'' spectrum was obtained while the system was 1.5 mag fainter than during previous observations), an indication that the flux normalization is not largely in error.  In AS 325, the flux differences may be explained by photometric variability associated with the system (though we do not have photometric observations of this star to confirm this) or may be associated with variable seeing, transparency, and sky conditions.  It should also be noted that AS 325 exhibits strong variability in its emission line characteristics possibly associated with an increased mass loss rate (see \S 3.7) which could, in turn, result in a drop in the continuum level seen in the spectra.   While the reason for the flux discrepancies may be astrophysical or observational, none of the results presented here are dependent on the absolute normalization of the spectra and thus it does not affect our analysis. 

The  metal emission lines, hydrogen emission lines, and Balmer absorption lines all occur at different velocities in the spectra of the two stars.  Table 3 quantifies the velocities for each species in both systems.  The velocities of each of the strong features of a given species are averaged to construct the ensemble velocity of that species.  The reported values are the average velocities measured from data spanning several nights of observations; typical dispersion around the mean was 80 \kms, giving an estimate of the average error in our velocity determinations. As noted by the referee, averaging velocities measured on lines of very different strengths may lead to a biased estimate of the average velocity as less opaque lines may be formed in different layers of the wind compared to more optically thick ones. To minimize this effect, we only use the four strongest Balmer lines as an estimate of the hydrogen emission line velocity.  For the metal species, which also exhibit lines of various strengths, we only use the brightest 20\% of the lines to determine the average velocity.  The velocity structure for both systems is similar; the absorption lines are the most blueshifted with the hydrogen emission lines, metal lines, and  \CaII\, and \OI lines being shifted progressively less to the blue.

\subsection{Environment}

\begin{figure}[hb]
\epsscale{1.3}
\plotone{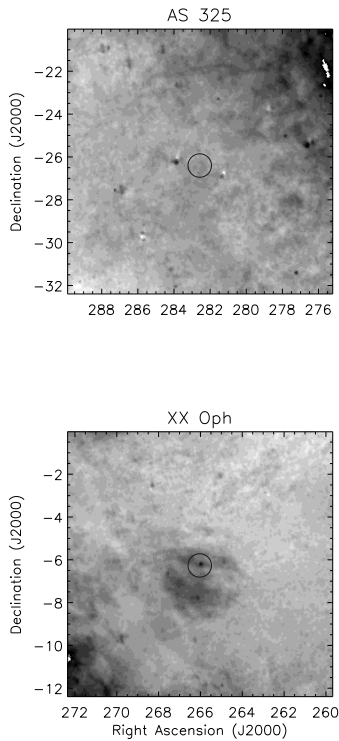}
\caption{ \scriptsize The environment of both ``iron'' stars from the Wisconsin Hydrogen Alpha Mapper. The bottom panel shows the \ha \, emission near XX Oph while the top panel illustrates the emission near AS 325.  Both stars (marked by the circles) are located in regions of significant H$\alpha$ emission.  Note that XX Oph is located particularly near a localized enhancement while AS 325 is not visible in this short WHAM exposure.}
\label{fig:wham}
\end{figure}

 Figure \ref{fig:wham} illustrates the environment of each star from the Wisconsin Hydrogen Alpha Mapper (WHAM) \citep{hal2003}. Both of these stars appear to reside in regions of enhanced hydrogen emission in our galaxy. The WHAM kinematic data show the \ha\, emission toward XX Oph (AS 325) occurs at $2.4 \pm 11.8$ ($17.5\pm 27.02$) \kms.  While the velocities of the interstellar gas are inconsistent with the Balmer emission line velocities measured in the  two ``iron'' stars, the velocities of the metals are  consistent within the measurement errors. It is possible that the emission signatures in these two stars are not localized near the stars but, instead, originate in extended regions along our line of site to them (i.e. arising from a phenomena associated with the ISM alone).  In order to probe this possibility, we obtained spectra of several similarly bright stars in the vicinity of the two ``iron'' stars of interest.  Table 4 lists the properties of the stars chosen for these observations.    None of these stars show any sign of the emission features noted in the spectra of XX Oph and AS 325.

The absence of emission line features in neighboring stars may simply indicate that stars chosen for comparison are all projected in front of the origin of the peculiar emission. If the emission features we find in both XX Oph and AS 325 arise in an  extended nebular region, a deep spectrum of nearby blank sky would reveal emission features similar to those seen in each of the ``iron'' stars.  A 1200s exposure of the sky near XX Oph obtained at the Bok 2.3m telescope on Kitt Peak revealed no emission due to metal lines. This indicates that the cause of the unusual spectra of AS 325 and XX Oph is intrinsic to the two systems and does not simply arise from unrelated phenomena along our line of sight to the stars.

\begin{deluxetable}{ccccc}[!ht]
\tablecolumns{3}
\tablenum{3}
\tabletypesize{\scriptsize}    
\tablewidth{0pt}
\tablecaption{Velocities for Several Species in XX Oph and AS 325}
\tablehead{   
\colhead{Species} &
\colhead{XX Oph} &
\colhead{AS 325}  \\
\colhead{} &
\colhead{km s$^{-1}$} &
\colhead{km s$^{-1}$}}
\startdata
H (Balmer) emission & -94 & -65  \\
H (Balmer) absorption & -280 & -227  \\
\hbox{He  \sc  i} emission & 0 & \nodata  \\
\hbox{He  \sc  i} absorption & -110 & \nodata  \\
\hbox{O  \sc  i} emission & 31 & 84  \\
\hbox{Ca  \sc  ii} emission & 85 & 90  \\
\hbox{Ca  \sc  ii H+K} absorption & \nodata & -77  \\
\hbox{Ti  \sc  ii} emission & -28 & -2  \\
\hbox{Cr  \sc  ii} emission & -31 & 2  \\
\hbox{Fe  \sc  ii} emission & -21 & -8  \\
\hbox{Na  \sc  i} absorption & \nodata & -68 
\enddata

\end{deluxetable}

\begin{deluxetable*}{ccccc}
\tablecolumns{5}
\tablenum{4}
\tablewidth{0pt}
\tabletypesize{\scriptsize}    
\tablecaption{Observed Comparison Stars}
\tablehead{
  \colhead{Star} & 
  \colhead{Right Ascension} &
  \colhead{Declination} & 
  \colhead{$\mbox{m}_V$\tablenotemark{1}} &
  \colhead{Observatory}  \\
  \colhead{} & 
  \colhead{(J2000)} & 
  \colhead{(J2000)} & 
  \colhead{(mag)} &
  \colhead{}}
\startdata
BD-07-4467  &    17:36:54.58 & -07:41:52.4 & 9.3 & Coud\'{e}  \\
BD-07-4473  &    17:38:53.24 & -07:44:24.2  & 9.5 & Coud\'{e}  \\
BD-07-4476   &   17:40:04.47  &-07:06:21.5  & 9.3 & Coud\'{e}  \\
NSV-9567   &  17:43:08.01  & -06:20:25.0  & 8.83  & Bok   \\
BD-06-4636 &  17:43:29.85  & -06:08:01.7 & 9.8  & Bok   \\
BD-05-4487 &  17:43:45.23  & -06:03:59.9  & 9.5   & Bok   \\
BD-06-4642 &  17:44:26.54  & -06:24:22.2  & 9.8    & Bok   \\
BD-07-4494 &  17:44:39.40  & -07:48:47.4  & 9.4  & Bok   \\
BD-07-4504 &     17:48:11.49 & -07:05:24.1 & 9.5 & Coud\'{e}  \\
PPM-734398 &     18:46:19.77 & -27:54:57.7 & 9.5& Coud\'{e}  \\
CPD-27-6538 & 18:50:21.15  & -26:55:25.0  & 9.5  & Bok   \\
CD-27-13276 & 18:50:44.23 &  -27:00:31.6 & 9.90  & Bok   \\
CD-26-13591 &  18:54:53.66  & -26:20:46.4 &  9.5  & Bok   \\
V4061-Sg  &      18:56:27.7  & -27:30:22 & 9.22 & Coud\'{e}  \\
\enddata
\tablenotetext{1}{All magnitudes were taken from the SIMBAD Astronomical Database}
\end{deluxetable*}

\subsection{Spectral Types}

\begin{figure*}[hb]
\centering{\includegraphics[angle=90, width=5.5in]{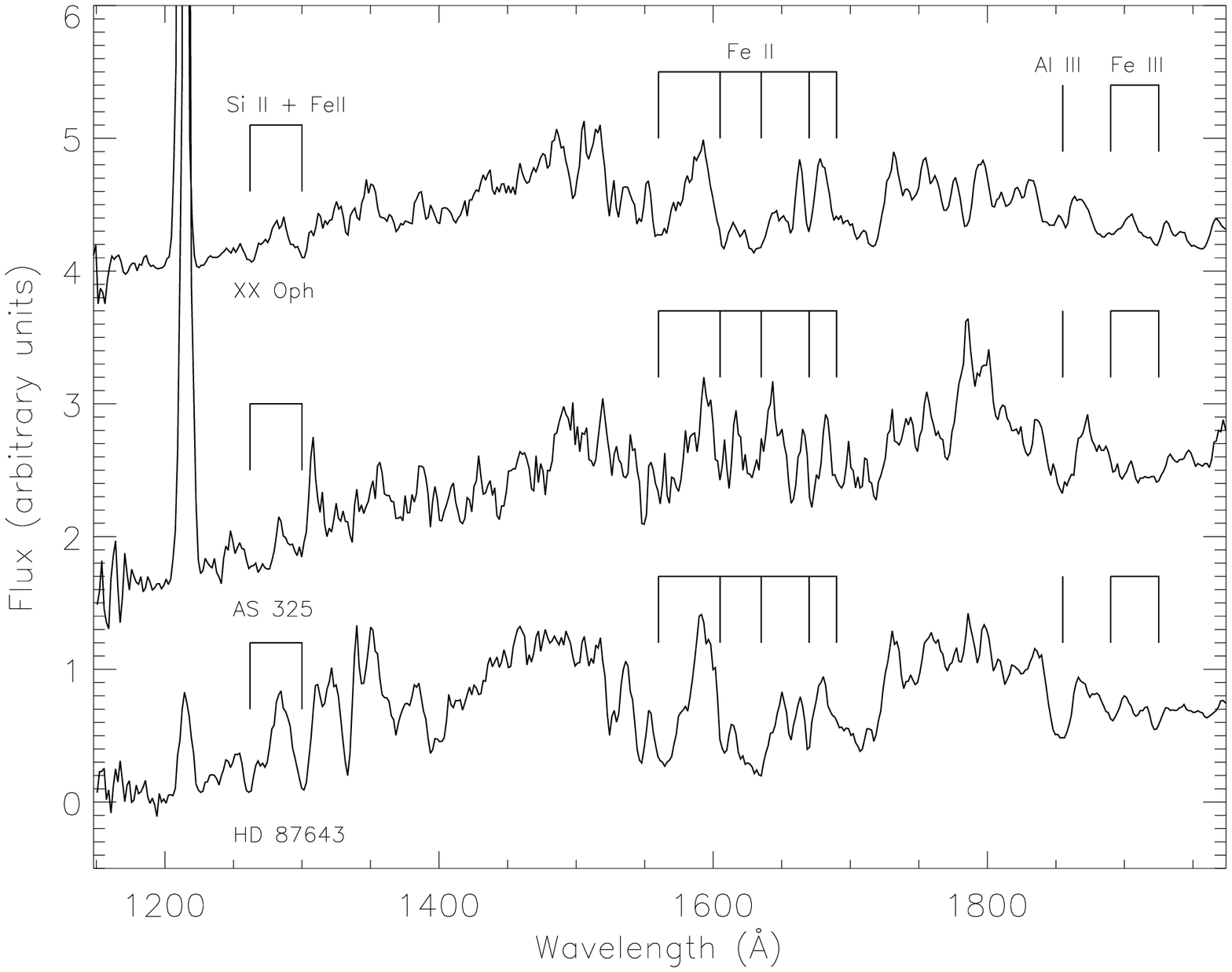}}
\caption{ \scriptsize Archival far-ultraviolet spectra of XX Oph, AS 325, and the Be star HD 87643 from IUE.  The three spectra are quite similar - all show deep absorption from \AlIII, a sign of a strong wind, as well as absorption from ionized iron.  The optical \FeII\, emission is likely pumped by these iron absorption lines through fluorescence.}
\label{fig:uvspec}
\end{figure*}

The optical spectra of both XX Oph and AS 325 are quite complex and do not resemble typical spectra, making any precise determination of the spectral type of the underlying stars difficult.  Observations at other wavelengths, however, can help explain the nature of these peculiar systems.  It has been noted previously that the far-ultraviolet spectrum of XX Oph shows strong absorption due to \FeII\, \citep{mich1991}.  The presence of strong ultraviolet flux from XX Oph indicates that it must have a hot component, suggested to be a Be star in other analyses \citep{ldr1975,dt1990}.  In the optical, the high reddening,  $A_V = 1.6-4.0$ depending on the analysis \citep{eal1993, ldr1975}, obscures the blue continuum, hiding the star.  Figure \ref{fig:uvspec} shows the ultraviolet spectra of XX Oph, AS 325, and HD 87643, a Be star \citep{pacheco}, taken from the IUE archive.  The three spectra look quite similar - all have strong signs of absorption due to \FeII\, and \AlIII\, with weaker absorption from \FeIII.  The similarity between the three spectra indicates that the stars are likely quite similar; XX Oph and AS 325 {\it both} likely have Be components.  

xxxxx 

\begin{figure*}[ht]
\centering{\includegraphics[angle=90, width=5in]{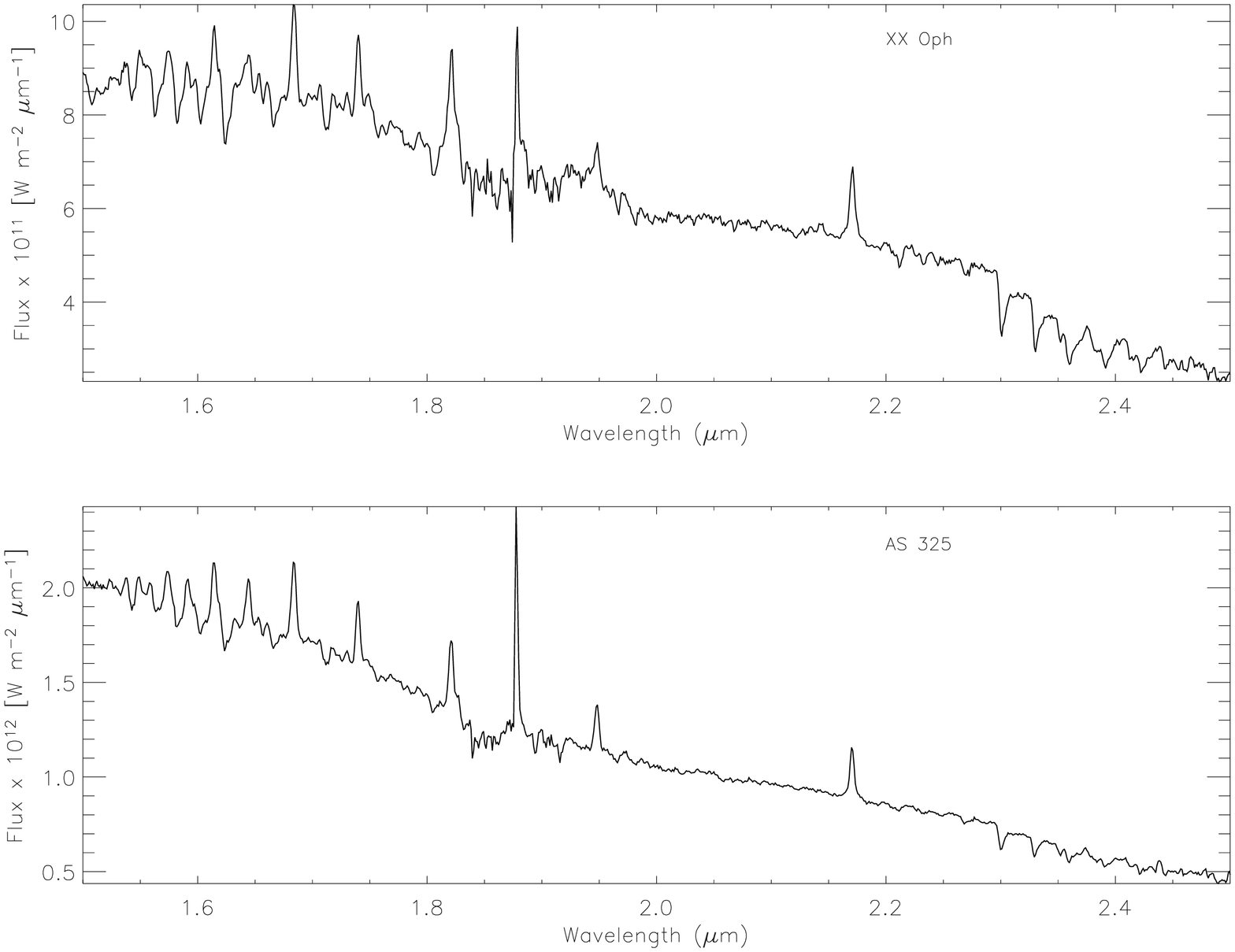}}
\caption{ \scriptsize Near infrared spectra of XX Oph and AS 325 obtained on 2003 Oct 10.  The spectral structure of the stars changes across this wavelength range; in the blue, the spectra are dominated by emission lines while the red end of the spectra is dominated by absorption signatures from the photosphere of the cool star in each system.}
\label{fig:irspec}
\end{figure*}

\begin{figure}[ht]
\centering{\includegraphics[angle=90, width=3.7in]{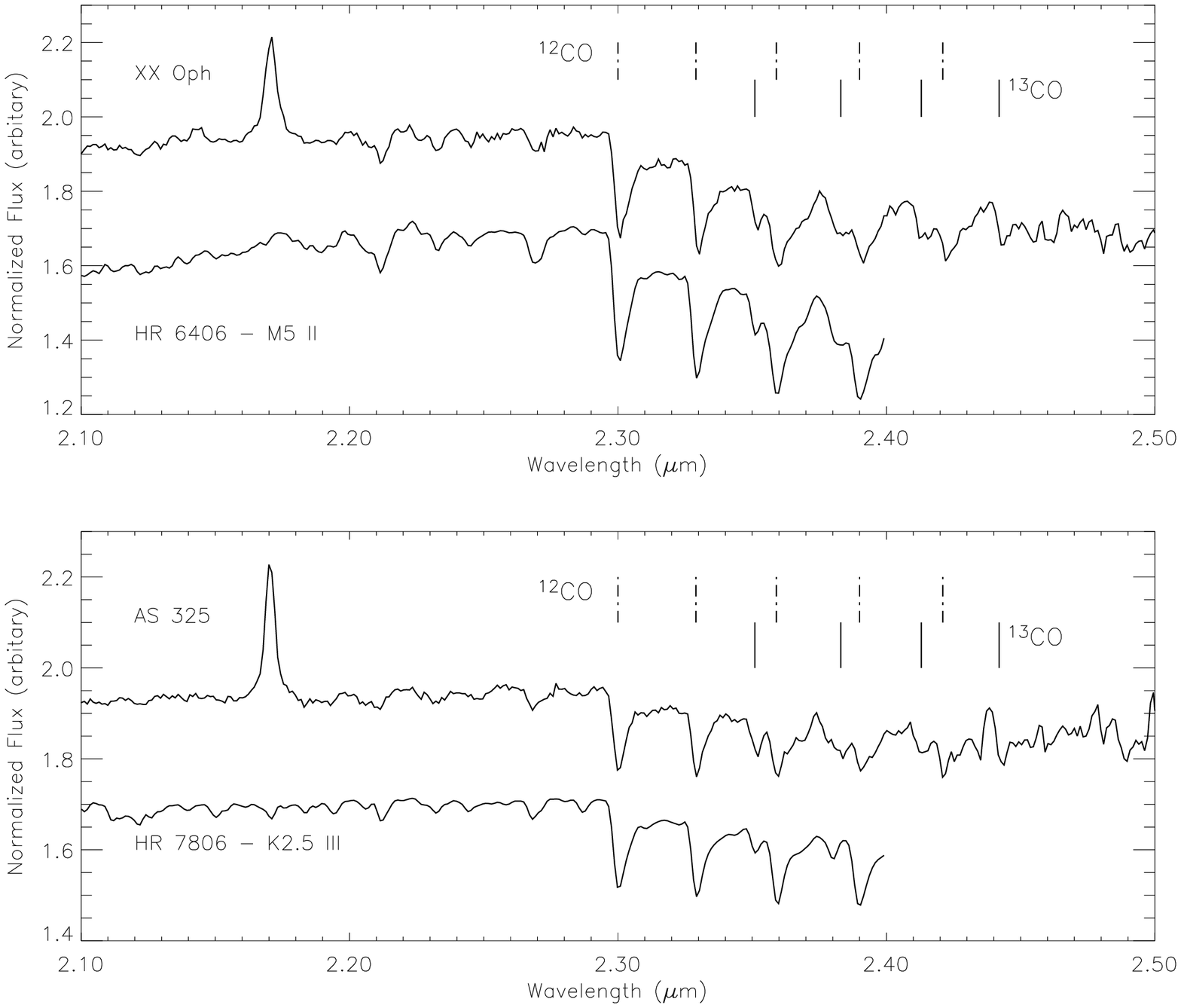}}
\caption{ \scriptsize  The CO region of both stars in detail.  For comparison, we show a spectral standard which fits the data well.  We find that the spectrum of XX Oph is well matched by an M5 II star, in close agreement with past analysis in the literature.  AS 325 resembles a K2.5 III star.  The two spectra in each panel are offset for clarity.}
\label{fig:irspecco}
\end{figure}

Our near infrared K-band spectra of these two stars are shown in Figure \ref{fig:irspec} while Figure \ref{fig:irspecco} shows the CO region of both stars in greater detail.  While a detailed analysis of the infrared features of these stars will be presented in a later paper (including near infrared integral field spectroscopy from 1.2-5 \micron\,  of the two stars and a previously unpublished ISO spectrum), these K-band spectra reveal the underlying stellar properties of the red stars in the systems and deserve mention here. The spectra are dominated by emission lines from the Brackett series and Paschen $\alpha$ in their blue regions.   Redward of 2.2 \micron \, the stellar photosphere finally reveals itself,  showing absorption due of $^{12}$CO, \NaI, \CaI,  and $^{13}$CO, a signature of evolved late-type stars.  We compare our K-band spectra with the stellar atlas of \citet{wh1997} to determine the approximate spectral type of the cool component in each system. While the spectra in this catalog do not cover all spectral types and luminosity classes, a fair evaluation of the spectral type of each star can be made. We find the K-band spectrum of XX Oph is well described by an M5 II star, in close agreement with past studies which have suggested an M6 III secondary \citep{ldr1975,hg1977}. AS 325 is a close match to that of HR 7806, a K2.5 III star.  Figure \ref{fig:irspecco} shows the CO region of both stars and the best fit spectrum for comparison.  The bottom panels of  Figures \ref{fig:xxophspec} and \ref{fig:as325spec} show stars with similar spectral type as that determined from our infrared spectra taken from the UVES library of high resolution field star spectra \citep{bagnulo} and convolved to our resolution.  In both  stars, the emission lines and the presence of the hot component contaminate the optical spectrum, making a direct comparison difficult, but there is good correspondence between the continuum shape of the comparison star spectra and that of the ``iron'' stars.

 Under the assumption that our infrared spectral classifications are correct, we can use the K-band magnitude of XX Oph from \citet{ka1999} and AS 325 from 2MASS to determine the distance to the systems.  We find that XX Oph (AS 325) is located at approximately 912 pc (1100 pc) if the cool component has the same properties as RV Hya (HR 7806) a known M5II (K2.5 III) star with a robust parallax.  This distance to XX Oph is within the 1-$\sigma$ error of that determined from a Hipparcos  parallax, $546^{+805}_{-205}$ pc.

\subsection{Stellar Radius}

We measured the angular size of XX Oph using the
Palomar Testbed Interferometer \citep{colavita1999}
in both H and K bands on 5 separate nights in 2003 using
the North-West baseline (85m). The visibility data were
calibrated using the standard method by \citet{boden1998},
utilizing three unresolved stars as calibrators: HD 161868
($0.73 \pm 0.10$ mas), HDC158352 ($0.60 \pm 0.10$ mas) and HD 173417
($0.60 \pm 0.10$ mas). The visibility data were then fit using
the uniform disk (UD) model, such that
\begin{equation}
V^2 =  \left(\frac{2 J_1\left(\pi B \theta / \lambda\right)}{\pi B \theta / \lambda}\right) ^2
\end{equation}
where $B$ is the projected baseline (m), $\lambda$ is the
wavelength of observation (m), $\theta$ is the UD angular
size (rad) and $J_1$ is the first-order Bessel function.

The mean angular diameter in the H-band is $1.83 \pm 0.06$ mas,
and in the K-band is $1.94 \pm 0.06$ mas. Assuming a nominal
20\% error in the distance of 912 pc, the stellar radii
in the two bands are $179\pm36 R_\sun$ and $190\pm38 R_\sun$,
respectively.  To estimate the bolometric flux from the cool component in XX Oph, the component for which we measure the diameter, we fit available photometry spanning the wavelength range 5550 \ang\, - 5 \micron\, \citep{allen1973, swings1972, gezari1993, ldr1975} with a blackbody spectrum; we find $F_{\rm bol}=10.2 \pm 1.5 \times 10^{-8}$ erg/cm$^2$/sec. We calculate the effective temperature using 
\begin{equation}
T_{{\rm eff}} = 2341 \left( \frac{F_{\rm bol}}{\phi^2} \right)^{1/4}
\end{equation}
where $F_{{\rm bol}}$ is the bolometric flux in units of $10^{-8}$ erg/cm$^2$/sec and $\phi$ is the measured angular diameter in mas.
This calculation yields effective temperatures 
in the H and K-bands of $3092\pm126$ K and $3001\pm118$ K. The derived radius and effective temperature of XX Oph are consistent with that of a
late-type (M6-M8) luminosity class II star \citep{vbelle1999}, in good agreement with the spectral type we determine from our K-band spectra. It should be noted that the spectral energy distribution of XX Oph is composed of both a hot and cool component, and thus any determination of the bolometric flux from one of the two stars may be contaminated by the presence of the second component.  Here, we use the bolometric flux only for an estimate of the effective temperature, and not for any in-depth analyses.

\subsection{Binarity}

In order to place limits on the possible binary nature of these two systems, we examined the hydrogen emission line velocities on both long and short timescales using a combination of the strong \ha\, line as well as several of the strongest Paschen lines present in the spectra.  Since the hydrogen emission  lines arise from the wind arising from the hot Be star component, these emission lines can be used as a proxy to trace the kinematics of the hot star and its radial velocity motion. Since these emission lines are not tied directly to the stellar photosphere but instead originate from the hot wind of the star, these emission lines are only an indirect tracer of the stellar kinematics.  On the longest timescales (months to years), there is no sign of coherent velocity changes in the two systems.  While the time sampling of our measurements is not systematic, it is still clear that there is no long-term gross radial velocity motion. On  2003 July 8, several observations of each star were obtained covering the red portion of the spectrum to look for short term variations in the velocities of the emission lines. Spectra of XX Oph and AS 325 were obtained as alternate pairs of 300 second exposures for over three hours.  The velocities of multiple hydrogen emission lines were found to show no measurable variation during the three hour period.  Given our resolution of approximately 2 \ang \, at 6500 \ang\,, corresponding to 92 \kms, we can centroid \ha, which is observed at signal-to-noise greater than 100, to a level of approximately 10 \kms (the ability to centroid goes as the resolution over the square root of the signal-to-noise ratio in the line). Allowing for an additional 20 \kms uncertainty induced by the asymmetric shape of the \ha\, line and possible wavelength calibration error, we place upper limits of 25 \kms\, on the maximum allowed radial velocity which could be induced by a binary companion. The lack of any gross radial velocity motions on short scales rules out the presence of an massive, previously unknown, close component not directly inferred from the spectra.  In order to place more stringent constraints on the long term radial motions of the stars,  time resolved K-band echelle spectroscopy, possible with current instrumentation at UKIRT, is needed.

\subsection{Origin of Spectral Features}

The ultraviolet  spectra observed by IUE are compelling evidence that both stars have a strongly reddened hot component. The optical and ultraviolet spectral features likely arise in a strong, radiation driven, wind around each system.  Both of the stars have wide and strong \AlIII\, absorption in the ultraviolet, probably formed in this  wind. For comparison, HD 87643 (see Figure \ref{fig:uvspec}), which shows a very similar ultraviolet spectrum to the two ``iron'' stars studied here, has a mass loss rate of $7\times10^{-7} M_\sun \mbox{yr}^{-1}$ \citep{pacheco}.  It is interesting to note that the velocity of the \AlIII\, absorption line in XX Oph (AS 325)  of $-240$ \kms\, ($-80$ \kms) agrees reasonably well with the Balmer absorption velocity of $-280$ \kms\, ($-227$ \kms) though the spectral resolution of IUE ($\sim 1000$ \kms) makes interpretation of this similarity difficult. Dense material in the expanding wind provides the P-Cygni absorption which is observed in the optically thin Balmer lines.  The ultraviolet spectrum, with strong signatures of \FeII\, absorption, is reminiscent of the ``iron curtain''  observed in novae eruptions \citep{hws1992, sal1994}.   The narrow optical emission lines of iron and other metals are pumped by  ultraviolet absorption lines through fluorescence \citep{sa1993}. In AS 325, the \NaI\, absorption line and the \CaII\, H+K lines  are likely residual absorption from the the cool K giant component in that system. While most of the optical absorption lines from the cool star are erased by the hot continuum from the Be star, these lines are quite strong in early K giants and thus leave a residual signature in AS 325.  These line may also arise in the ISM, but the agreement between the observed Balmer line velocity and the \NaI\, and \CaII\, velocities, combined with the rather large equivalent widths of these lines (1.9 and 1.2 \ang respectively for the stronger line in each doublet) argue against this interpretation.

\begin{figure*}[ht]
\centering{\includegraphics[angle=90, width=7in]{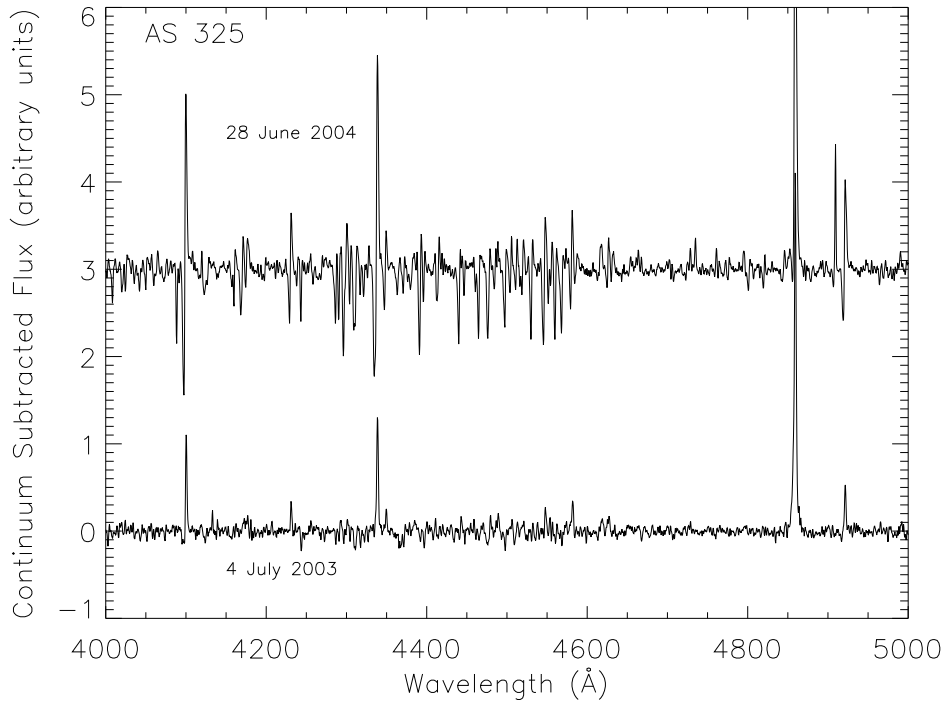}}
\caption{ \scriptsize Spectral variability of the star AS 325 over a one year baseline.  The bottom spectrum is AS 325 on 4 July 2003 while the upper spectrum (offset vertically for clarity) shows AS 325 on 28 June 2004.   The differences are striking.  The P-Cygni profiles in the Balmer lines, which are present but weak in the earlier spectrum, become quite strong by the upper spectrum.  The \FeII\, lines show strong P-Cygni profiles in the top spectrum, as well.  This evolution is interpreted as indicating density changes in the Be star wind. }
\label{fig:compareas325}
\end{figure*}

\subsection{Variability}

Given the large amount of optical spectroscopy we have for these two stars, it is a natural extension to examine the spectral variability of each.  While the gross properties of XX Oph remained constant throughout our observational campaign, AS 325 showed peculiar spectral signatures in 2004.  Figure \ref{fig:compareas325} illustrates this difference; the Balmer absorption components become quite strong and the \FeII\, lines show P-Cygni profiles in our newest spectrum.  The appearance in the optical \FeII\, absorption features is likely connected to density variations in the wind.  If the mass loss rate of the hot star changes, adding more material to the wind, or if the wind begins to sweep up more material from the ambient ISM, the density of the surrounding gas will increase, leading to stronger absorption components. 

 \citet{bh1989} presented a spectrum of AS 325 near the \NaI\, D absorption lines which is quite different to the spectrum presented in Figure \ref{fig:as325spec}.  In 1987, AS 325 was observed to have  strong \NaI\, D absorption lines, similar to those observed in Figure \ref{fig:as325spec}, but these absorption lines were accompanied by slightly redshifted emission peaks.  The \NaI\, D lines did not show a P-Cygni profile, but instead appear to be the superposition of two independent components.  The \NaI\, D  emission lines are not present in any of our recent observations.  Their disappearance may be related to changes in the photo-ionization background generated from the central hot star which in turn changes the ionization state of low ionization species such as \NaI.

\begin{figure}[ht]
\centering{\includegraphics[angle=90, width=3.5in]{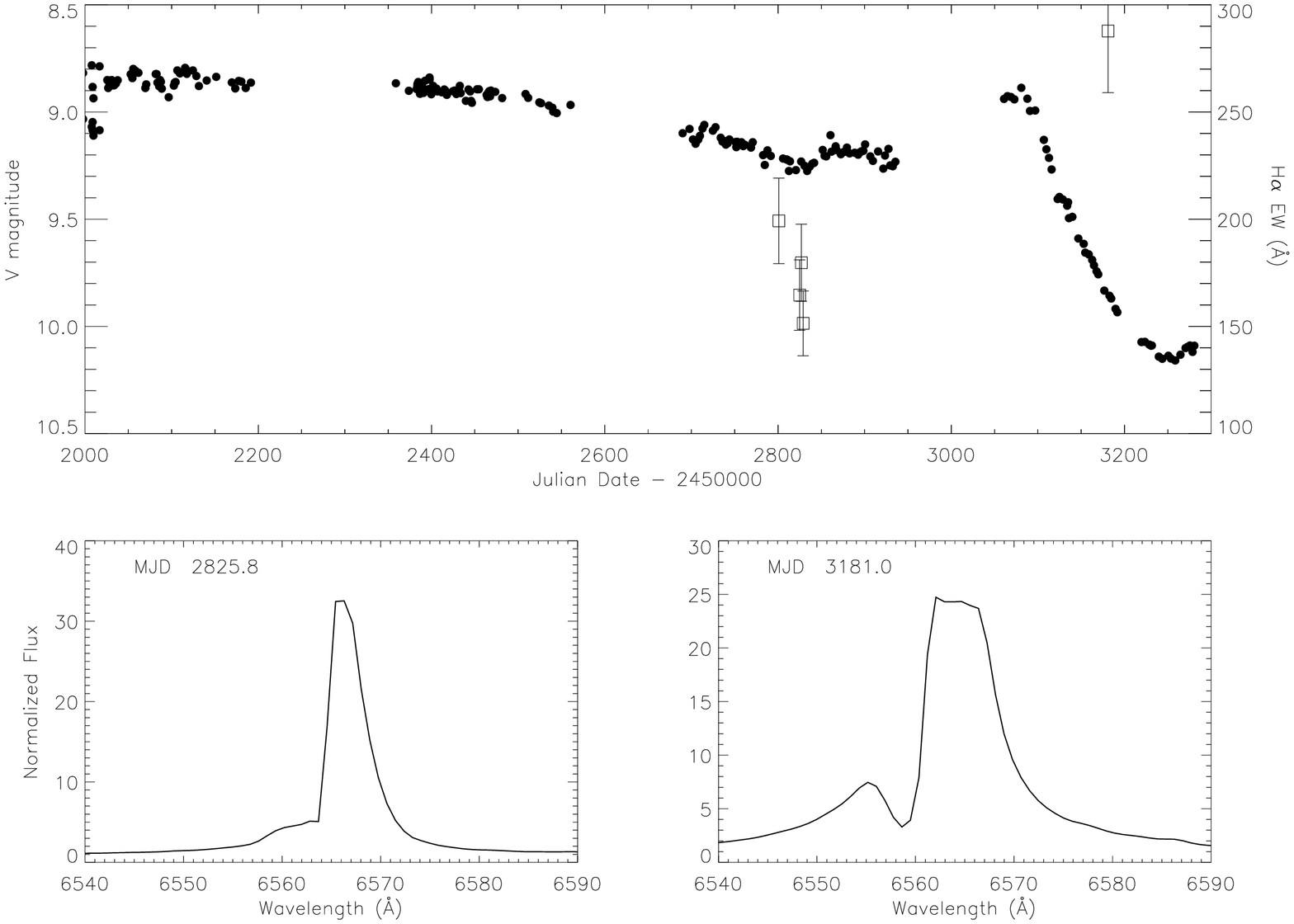}}
\caption{ \scriptsize The spectral and photometric variability of XX Oph.  The top panel shows the ASAS-3 V magnitude of XX Oph from \citet{sobotka2004} in filled points; the empty squares show the \ha\, equivalent width measured from our spectra.  When XX Oph is in its faint state, the \ha\, equivalent width increases by nearly 50\%.  The bottom panels illustrate the \ha\, line morphology both before and during this photometric event.  While XX Oph is in its faint state, the absorption component of the P-Cygni profile becomes apparent while in its normal state the \ha\, line is merely asymmetric.  Although the peak flux in less while the star is in its faint state, the width of the line is larger, increasing the equivalent width of the line.  The change in the \ha\, profile and equivalent width are likely due to variability in the late-type component in the system.}
\label{fig:variability}
\end{figure}

In March 2004, the optical flux decreased by 1.5 magnitudes in XX Oph, the deepest minimum in 37 years \citep{sobotka2004}.  Figure \ref{fig:variability} shows the variations in the \ha\, equivalent width compared to the ASAS-3 light curve presented by \citet{sobotka2004} and  the variation in the \ha\, line shape both before and during the photometric event.  In its faint state,  the \ha\, equivalent width is 50\% larger than before the event and the presence of the P-Cygni absorption component is more pronounced after the onset of the photometric event.  The appearance of the absorption component and the increased equivalent width indicates that the underlying continuum has faded from its normal state. Our spectra show that the continuum flux at \ha\, decreased by a factor of 6 between 5 July 2003 and 25 June 2004. The lower continuum provides less flux to contaminate the absorption component, allowing the P-Cygni profile to become clear.  The change in both the photometric flux and spectroscopic features requires dramatic variability; since evolved late-type stars tend to be highly variable, for example through pulsarations or  a dynamic mass loss rate, we suspect the continuum variability is associated with late-type star.  The smooth decrease in optical flux shown in Figure \ref{fig:variability} is remaniscent of an eclipse. While this could explain the decrease in the flux needed to produce the spectral variations, historical light curves of the system show that the system undergoes sporadic deep minima, often lasting years, remaniscent of R Coronae Borealis stars \citep{prager}. Althought the photometric propetries of XX Oph are similar to those of R Coronae Borealis stars, our spectra of XX Oph do not contain any of the spectral features characteristic of this class, particularly the strong \CII\, Swan bands which appear in deep minima.

\subsection{Longevity}

It is interesting to note the similarities between the model presented here and that invoked to describe emission from nova eruptions \citep{Gaposchkin1957, williams1992}.  It is natural to ask, therefore, if the emission observed in XX Oph and AS 325 could be described by the nova model.  The key characteristics of the spectrum of XX Oph have remained constant since Merrill's work in the first half of the twentieth century.  In contrast, the nova phase with similar emission features lasts only a few weeks. Thus, the nova interpretation cannot describe the longevity of the observed  features in XX Ophiuchi and AS 325.

Using his photographic  spectra, \citet{merrill1951}, measured the velocities of the metal and hydrogen emission features to be nearly the same as those presented here.  In 1951, however, the absorption component of the hydrogen lines was measured to be approximately $-60$ \kms, nearly a factor of four slower than we measure more recently.   From the velocity differences between 1951 and 2004, it appears the structure of the outflowing wind from XX Oph has changed between these two observational epochs. In order to create the measured velocity shifts, either the maximum speed of the wind or the density in the wind must have increased. An increased mass loss rate could also explain the photometric variation seen in XX Oph; if the mass loss rate increases, the system may become more heavily obscured, resulting in a drop in visible flux.  Future high resolution spectroscopic monitoring of both stars are needed to understand the systematics of the wind associated with this system.

 \section{Conclusions}

We have obtained new optical and infrared spectra as well as archival far ultraviolet spectra of the two ``iron'' stars XX Oph and AS 325.   The two stars are quite similar, contrary to claims in the literature, indicating that these two systems are examples of one class of object.  Our new data indicate that these stars are both binary systems with a hot Be component as well as an evolved late-type star.  The forest of \FeII\, lines observed in the optical are likely pumped by ultraviolet \FeII\, absorption lines through fluorescence, and the Balmer P-Cygni profiles arise in the expanding wind around each system.  We find that the velocity of the absorption component of the Balmer lines in XX Oph is larger than that measured in the past, evidence that the stellar wind in XX Oph has accelerated over the past fifty years - a sign that these stars are dynamic and possibly observed in a short lived phase. Also, we examine the spectral variability of XX Oph during a deep photometric fading and find that the shape and strength of the \ha\, line changes quite strongly during the event.   Further follow-up spectroscopic observations are vital in order to understand the connection between the peculiar photometric and spectral variability in these stars.    In order to probe the stellar properties of XX Oph and AS 325, infrared spectra are needed to penetrate the forest of iron lines and the continuum from the hot component which confuses the observed continuum in the optical.  In a future paper, we will present near infrared integral field spectra of these two stars and an archival ISO spectrum which more directly probe both XX Oph and AS 325.

\section{Acknowledgments}

We are grateful to Kristoffer Eriksen who provided many enlightening conversations about early drafts of this paper.  We also thank our anonymous referee for an in-depth, critical, look at our work which resulted in a large improvement to our analysis.  John Moustakas kindly provided his IDL code, {\tt iSPEC}, for the reduction of the spectra gathered at the Bok 2.3 meter telescope. R. Cool was funded through a National Science Foundation Graduate Research Fellowship. 

The authors thank the director of Kitt Peak National Observatory, Richard Green, for his continued support of the Coud\'{e} Feed telescope and Daryl Willmarth for his masterful work at instrument setup.  Some of the data used in this research were collected by the NSF funded NOAO Teacher Leader Research Based Science Education (TLRBSE) Project.   The TLRBSE Project is funded by the National Science Foundation under ESI 0101982, funded through the AURA/NSF Cooperative Agreement AST-9613615.  NOAO is operated by the Association of Universities of Research in Astronomy (AURA), Inc. under cooperative agreement with the National Science Foundation.  The United Kingdom Infrared Telescope is operated by the Joint Astronomy Centre on behalf of the U.K. Particle Physics and Astronomy Research Council. Observations at PTI are made possible through the efforts of
the PTI Collaboration, and part of this work was performed
at the Jet Propulsion Laboratory under contract with the
National Aeronautics and Space Administration.

Some of the data presented in this paper were obtained from the Multimission Archive at the Space Telescope Science Institute (MAST). STScI is operated by the Association of Universities for Research in Astronomy, Inc., under NASA contract NAS5-26555. Support for MAST for non-HST data is provided by the NASA Office of Space Science via grant NAG5-7584 and by other grants and contracts.  The Wisconsin H-Alpha Mapper is funded by the National
Science Foundation. The NIST Atomic Spectra Database was used to determine line identifications for this work.  The UVES Paranal Observatory Project (ESO DDT ProgramID 266.D-5655) provided the optical comparison spectra presented here.  This publication makes use of data products from the Two Micron All Sky Survey, which is a joint project of the University of Massachusetts and the Infrared Processing and Analysis Center/California Institute of Technology, funded by the National Aeronautics and Space Administration and the National Science Foundation.This research has also made use of the SIMBAD database, operated at CDS, Strasbourg, France.

\begin{deluxetable*}{cccccccccc}
\tablecolumns{10}
\tablewidth{0pt}
\tabletypesize{\scriptsize}
\tablenum{2}
\tablecaption{Prominent Spectral Lines in XX Oph and AS 325}
\tablehead{   
\colhead{$\lambda_{\mbox{lab}}$} &
\colhead{Species} &
\colhead{Line Type} & 
\colhead{$\lambda_{\mbox{obs}}$} & 
\colhead{$\lambda_{\mbox{obs}}$} &
\colhead{$\lambda_{\mbox{lab}}$} &
\colhead{Species} &
\colhead{Line Type} & 
\colhead{$\lambda_{\mbox{obs}}$} & 
\colhead{$\lambda_{\mbox{obs}}$} \\
\colhead{\ang} &
\colhead{} &
\colhead{} &
\colhead{XX Oph}&
\colhead{AS 325}&
\colhead{\ang}&
\colhead{}&
\colhead{}& 
\colhead{XX Oph}&
\colhead{AS325} \\
\colhead{(1)} &
\colhead{(2)} &
\colhead{(3)} &
\colhead{(4)} & 
\colhead{(5)} & 
\colhead{(6)} &
\colhead{(7)} & 
\colhead{(8)} & 
\colhead{(9)} &
\colhead{(10)}}
\startdata

  3771.74 &                    H &          A &    3768.30 &    3768.80   &      4296.57 &     \hbox{Fe \sc ii} &	   E &    4296.02 &    \nodata \\
   3771.74 &                    H &          E &    3770.39 &    3770.44  &      4300.05 &     \hbox{Ti \sc ii} &	   E &    4299.33 &    4299.76 \\
   3786.32 &     \hbox{Ti \sc ii} &          E &    3785.49 &    \nodata  &     4303.18 &     \hbox{Fe \sc ii} &   E &    4302.56 &    4303.12         \\
   3799.02 &                    H &          A &    3795.52 &    3795.63  &     4305.89 &     \hbox{Fe \sc ii} &	  E &	 4305.15 &    \nodata \\
   3799.02 &                    H &          E &    3798.23 &    3799.71  &     4307.86 &     \hbox{Ti \sc ii} &	  E &	 4307.21 &    4307.65 \\
   3814.12 &     \hbox{Fe \sc ii} &          E &    3813.37 &    \nodata  &     4320.96 &     \hbox{Ti \sc ii} &	  E &	 4320.21 &    4320.98 \\
   3819.60 &      \hbox{He \sc i} &          A &    3817.67 &    \nodata  &     4330.69 &     \hbox{Ti \sc ii} &	  E &	 4329.96 &    \nodata \\
   3819.60 &      \hbox{He \sc i} &          E &    3819.83 &    \nodata  &     4341.74 &		     H &	  A &	 4337.95 &    4338.75 \\
   3824.93 &     \hbox{Fe \sc ii} &          E &    3824.18 &    \nodata  &     4341.74 &		     H &	  E &	 4340.14 &    4340.59 \\
   3827.08 &     \hbox{Fe \sc ii} &          E &    3826.64 &    \nodata  &     4346.85 &     \hbox{Fe \sc ii} &	  E &	 4346.06 &    \nodata \\
   3833.70 &      \hbox{He \sc i} &          A &    3833.28 &    \nodata  &     4351.77 &     \hbox{Fe \sc ii} &	  E &	 4351.22 &    4351.56 \\
   3833.70 &      \hbox{He \sc i} &          E &    3834.21 &    \nodata  &     4359.33 &     \hbox{Fe \sc ii} &	  E &	 4358.63 &    \nodata \\
   3836.52 &                    H &          E &    3838.20 &    3836.82  &     4367.66 &     \hbox{Ti \sc ii} &	  E &	 4367.05 &    \nodata \\
   3888.60 &      \hbox{He \sc i} &          A &    3886.98 &    \nodata  &     4369.41 &     \hbox{Fe \sc ii} &	  E &	 4368.75 &    \nodata \\
   3888.60 &      \hbox{He \sc i} &          E &    3888.10 &    \nodata  &     4374.81 &     \hbox{Ti \sc ii} &	  E &	 4374.10 &    4375.50 \\
   3890.20 &                    H &          E &    3888.90 &    3889.64  &     4385.39 &     \hbox{Fe \sc ii} &	  E &	 4384.57 &    \nodata \\
   3900.55 &     \hbox{Ti \sc ii} &          E &    3900.12 &    \nodata  &     4391.03 &     \hbox{Ti \sc ii} &	  E &	 4390.48 &    \nodata \\
   3906.03 &     \hbox{Fe \sc ii} &          E &    3905.23 &    \nodata  &     4395.03 &     \hbox{Ti \sc ii} &	  E &	 4394.58 &    \nodata \\
   3914.50 &     \hbox{Fe \sc ii} &          E &    3913.92 &    3914.33  &     4399.77 &     \hbox{Ti \sc ii} &	  E &	 4399.47 &    \nodata \\
   3933.66 &     \hbox{Ca \sc ii} &          A &    \nodata &    3932.64  &     4411.07 &     \hbox{Ti \sc ii} &	  E &	 4410.51 &    \nodata \\
   3938.97 &     \hbox{Fe \sc ii} &          E &    3938.24 &    \nodata  &     4413.78 &     \hbox{Fe \sc ii} &	  E &	 4413.14 &    \nodata \\
   3968.47 &                  CaI &          A &    \nodata &    3967.62  &     4416.83 &     \hbox{Fe \sc ii} &	  E &	 4416.39 &    \nodata \\
   3971.24 &                    H &          A &    3967.07 &    3968.73  &     4432.45 &     \hbox{Fe \sc ii} &	  E &	 4431.78 &    4432.72 \\
   3971.24 &                    H &          E &    3970.06 &    3970.62  &     4443.79 &     \hbox{Ti \sc ii} &	  E &	 4443.46 &    \nodata \\
   3974.17 &     \hbox{Fe \sc ii} &          E &    3973.90 &    \nodata  &     4457.94 &     \hbox{Fe \sc ii} &	  E &	 4457.74 &    \nodata \\
   3987.60 &     \hbox{Ti \sc ii} &          E &    3987.21 &    \nodata  &     4464.45 &     \hbox{Ti \sc ii} &	  E &	 4463.97 &    4464.48 \\
   4025.13 &     \hbox{Ti \sc ii} &          E &    4024.82 &    \nodata  &     4468.51 &     \hbox{Fe \sc ii} &	  E &	 4468.18 &    \nodata \\
   4054.08 &     \hbox{Cr \sc ii} &          E &    4053.74 &    \nodata  &     4468.51 &     \hbox{Ti \sc ii} &	  E &	 4468.61 &    \nodata \\
   4102.94 &                    H &          A &    4099.79 &    4100.37  &     4472.93 &     \hbox{Fe \sc ii} &	  E &	 4472.49 &    4472.63 \\
   4102.94 &                    H &          E &    4101.71 &    4102.14  &     4489.18 &     \hbox{Fe \sc ii} &	  E &	 4488.70 &    4488.56 \\
   4122.67 &     \hbox{Fe \sc ii} &          E &    4122.43 &    \nodata  &     4491.40 &     \hbox{Fe \sc ii} &	  E &	 4490.90 &    4491.15 \\
   4124.79 &     \hbox{Fe \sc ii} &          E &    4124.59 &    \nodata  &     4501.27 &     \hbox{Ti \sc ii} &	  E &	 4500.76 &    \nodata \\
   4128.75 &     \hbox{Fe \sc ii} &          E &    4128.15 &    \nodata  &     4515.34 &     \hbox{Fe \sc ii} &	  E &	 4514.89 &    4514.98 \\
   4161.54 &     \hbox{Ti \sc ii} &          E &    4161.03 &    4161.51  &     4520.22 &     \hbox{Fe \sc ii} &	  E &	 4519.72 &    4520.00 \\
   4173.46 &     \hbox{Fe \sc ii} &          E &    4173.15 &    \nodata  &     4522.63 &     \hbox{Fe \sc ii} &	  E &	 4522.22 &    4522.65 \\
   4178.86 &     \hbox{Fe \sc ii} &          E &    4178.48 &    4178.88  &     4529.47 &     \hbox{Ti \sc ii} &	  E &	 4528.71 &    4529.13 \\
   4233.17 &     \hbox{Fe \sc ii} &          E &    4232.77 &    4232.97  &     4533.97 &     \hbox{Ti \sc ii} &	  E &	 4533.68 &    4534.23 \\
   4251.44 &     \hbox{Fe \sc ii} &          E &    4250.60 &    \nodata  &     4534.17 &     \hbox{Fe \sc ii} &	  E &	 4533.70 &    4534.69 \\
   4258.15 &     \hbox{Fe \sc ii} &          E &    4257.73 &    4257.88  &     4539.60 &     \hbox{Cr \sc ii} &	  E &	 4539.01 &    4539.88 \\
   4273.33 &     \hbox{Fe \sc ii} &          E &    4272.73 &    \nodata  &     4541.52 &     \hbox{Fe \sc ii} &	  E &	 4541.20 &    \nodata \\
   4287.39 &     \hbox{Fe \sc ii} &          E &    4287.01 &    \nodata  &     4549.47 &     \hbox{Fe \sc ii} &	  E &	 4549.07 &    4549.58 \\
   4287.87 &     \hbox{Ti \sc ii} &          E &    4287.71 &    \nodata  &     4549.62 &     \hbox{Ti \sc ii} &	  E &	 4549.61 &    4549.94 \\
   4290.22 &     \hbox{Ti \sc ii} &          E &    4289.70 &    4290.13  &     4555.89 &     \hbox{Fe \sc ii} &	  E &	 4555.44 &    4555.76 \\
   4294.10 &     \hbox{Ti \sc ii} &          E &    4293.68 &    4294.01  &     4558.65 &     \hbox{Cr \sc ii} &	  E &	 4558.18 &    4558.57 \\
\enddata
\end{deluxetable*}

\begin{deluxetable*}{cccccccccc}
\tablecolumns{10}
\tablewidth{0pt}
\tabletypesize{\scriptsize}
\tablenum{2}
\tablecaption{}
\tablehead{   
\colhead{$\lambda_{\mbox{lab}}$} &
\colhead{Species} &
\colhead{Line Type} & 
\colhead{$\lambda_{\mbox{obs}}$} & 
\colhead{$\lambda_{\mbox{obs}}$} &
\colhead{$\lambda_{\mbox{lab}}$} &
\colhead{Species} &
\colhead{Line Type} & 
\colhead{$\lambda_{\mbox{obs}}$} & 
\colhead{$\lambda_{\mbox{obs}}$} \\
\colhead{\ang} &
\colhead{} &
\colhead{} &
\colhead{XX Oph}&
\colhead{AS 325}&
\colhead{\ang}&
\colhead{}&
\colhead{}& 
\colhead{XX Oph}&
\colhead{AS325} \\
\colhead{(1)} &
\colhead{(2)} &
\colhead{(3)} &
\colhead{(4)} & 
\colhead{(5)} & 
\colhead{(6)} &
\colhead{(7)} & 
\colhead{(8)} & 
\colhead{(9)} &
\colhead{(10)}}
\startdata
 
  4563.76 &	\hbox{Ti \sc ii} &	    E &    4563.35 &	4563.77 &         4874.48 &	\hbox{Fe \sc ii} &	    E &    4874.21 &	\nodata \\
  4571.97 &	\hbox{Ti \sc ii} &	    E &    4571.40 &	\nodata &         4876.47 &	\hbox{Cr \sc ii} &	    E &    4875.93 &	\nodata \\
  4576.34 &	\hbox{Fe \sc ii} &	    E &    4575.93 &	4576.03 &         4884.61 &	\hbox{Cr \sc ii} &	    E &    4883.75 &	\nodata \\
  4583.84 &	\hbox{Fe \sc ii} &	    E &    4583.31 &	4583.83 &         4889.62 &	\hbox{Fe \sc ii} &	    E &    4889.01 &	\nodata \\
   4588.20 &     \hbox{Cr \sc ii} &          E &    4587.85 &    4588.09 &        4893.82 &	\hbox{Fe \sc ii} &	    E &    4893.29 &	\nodata \\
   4589.90 &     \hbox{Cr \sc ii} &          E &    4589.44 &    \nodata &        4911.19 &	\hbox{Ti \sc ii} &	    E &    4910.59 &	\nodata \\
   4589.96 &     \hbox{Ti \sc ii} &          E &    4589.39 &    \nodata &         4923.93 &	 \hbox{Fe \sc ii} &	   E &    4923.33 &    4923.53  \\
   4592.05 &     \hbox{Cr \sc ii} &          E &    4591.58 &    \nodata &         4950.74 &	 \hbox{Fe \sc ii} &	     E &    4950.25 &	 \nodata   \\
   4616.63 &     \hbox{Cr \sc ii} &          E &    4616.19 &    \nodata &         4990.50 &	 \hbox{Fe \sc ii} &	     E &    4990.19 &	 \nodata   \\
   4618.80 &     \hbox{Cr \sc ii} &          E &    4618.25 &    \nodata &         4993.36 &	 \hbox{Fe \sc ii} &	     E &    4992.89 &	 4993.40   \\
   4620.52 &     \hbox{Fe \sc ii} &          E &    4620.03 &    4620.68 &         5000.74 &	 \hbox{Fe \sc ii} &	     E &    5000.66 &	 \nodata   \\
   4629.34 &     \hbox{Fe \sc ii} &          E &    4628.85 &    4629.04 &         5005.51 &	 \hbox{Fe \sc ii} &	     E &    5005.31 &	 5005.09   \\
   4634.07 &     \hbox{Cr \sc ii} &          E &    4633.94 &    \nodata &         5018.44 &	 \hbox{Fe \sc ii} &	     E &    5018.02 &	 5018.23   \\
   4635.32 &     \hbox{Fe \sc ii} &          E &    4635.30 &    4634.21 &         5072.28 &	 \hbox{Ti \sc ii} &	     E &    5071.87 &	 \nodata   \\
   4639.67 &     \hbox{Fe \sc ii} &          E &    4639.04 &    \nodata &         5072.39 &	 \hbox{Fe \sc ii} &	     E &    5071.88 &	 \nodata   \\
   4656.98 &     \hbox{Fe \sc ii} &          E &    4656.47 &    \nodata &         5100.66 &	 \hbox{Fe \sc ii} &	     E &    5100.22 &	 5100.36   \\
   4657.21 &     \hbox{Ti \sc ii} &          E &    4656.47 &    \nodata &         5107.94 &	 \hbox{Fe \sc ii} &	     E &    5107.37 &	 \nodata   \\
   4666.76 &     \hbox{Fe \sc ii} &          E &    4666.27 &    \nodata &         5111.63 &	 \hbox{Fe \sc ii} &	     E &    5110.94 &	 5112.35   \\
   4670.18 &     \hbox{Fe \sc ii} &          E &    4669.62 &    \nodata &         5129.15 &	 \hbox{Ti \sc ii} &	     E &    5128.82 &	 \nodata   \\
   4697.60 &     \hbox{Cr \sc ii} &          E &    4697.03 &    \nodata &         5132.67 &	 \hbox{Fe \sc ii} &	     E &    5132.36 &	 5130.79   \\
   4708.67 &     \hbox{Ti \sc ii} &          E &    4708.29 &    \nodata &         5136.80 &	 \hbox{Fe \sc ii} &	     E &    5136.80 &	 \nodata   \\
   4713.16 &      \hbox{He \sc i} &          A &    4711.33 &    \nodata &         5154.07 &	 \hbox{Ti \sc ii} &	     E &    5154.19 &	 \nodata   \\
   4713.16 &      \hbox{He \sc i} &          E &    4712.91 &    \nodata &         5158.00 &	 \hbox{Fe \sc ii} &	     E &    5158.51 &	 \nodata   \\
   4720.15 &     \hbox{Fe \sc ii} &          E &    4719.22 &    \nodata &         5169.03 &	 \hbox{Fe \sc ii} &	     E &    5168.92 &	 5169.66   \\
   4728.07 &     \hbox{Fe \sc ii} &          E &    4727.57 &    4727.87 &         5172.47 &	 \hbox{Fe \sc ii} &	     E &    5172.28 &	 \nodata   \\
   4731.45 &     \hbox{Fe \sc ii} &          E &    4731.02 &    \nodata &         5184.79 &	 \hbox{Fe \sc ii} &	     E &    5183.60 &	 \nodata   \\
   4745.48 &     \hbox{Fe \sc ii} &          E &    4745.00 &    \nodata &         5188.68 &	 \hbox{Ti \sc ii} &	     E &    5188.39 &	 \nodata   \\
   4772.06 &     \hbox{Fe \sc ii} &          E &    4771.35 &    \nodata &         5197.58 &	 \hbox{Fe \sc ii} &	     E &    5197.63 &	 5197.63   \\
   4774.72 &     \hbox{Fe \sc ii} &          E &    4774.01 &    \nodata &         5226.54 &	 \hbox{Ti \sc ii} &	     E &    5226.56 &	 \nodata   \\
   4779.98 &     \hbox{Ti \sc ii} &          E &    4779.44 &    \nodata &         5227.49 &	 \hbox{Fe \sc ii} &	     E &    5227.91 &	 5227.10   \\
   4798.27 &     \hbox{Fe \sc ii} &          E &    4797.95 &    4798.57 &         5234.62 &	 \hbox{Fe \sc ii} &	     E &    5234.43 &	 5234.45   \\
   4798.52 &     \hbox{Ti \sc ii} &          E &    4798.00 &    \nodata &         5247.95 &	 \hbox{Fe \sc ii} &	     E &    5247.40 &	 \nodata   \\
   4805.08 &     \hbox{Ti \sc ii} &          E &    4804.51 &    \nodata &         5264.18 &	 \hbox{Fe \sc ii} &	     E &    5264.30 &	 5263.80   \\
   4812.34 &     \hbox{Cr \sc ii} &          E &    4811.75 &    \nodata &         5276.00 &	 \hbox{Fe \sc ii} &	     E &    5275.94 &	 5276.17   \\
   4814.53 &     \hbox{Fe \sc ii} &          E &    4813.74 &    \nodata &         5284.11 &	 \hbox{Fe \sc ii} &	     E &    5283.51 &	 5284.21   \\
   4824.13 &     \hbox{Cr \sc ii} &          E &    4823.76 &    \nodata &         5306.18 &	 \hbox{Fe \sc ii} &	     E &    5306.15 &	 5305.87   \\
   4833.20 &     \hbox{Fe \sc ii} &          E &    4832.57 &    \nodata &         5316.23 &	 \hbox{Fe \sc ii} &	     E &    5316.33 &	 5316.53   \\
   4836.23 &     \hbox{Cr \sc ii} &          E &    4835.60 &    \nodata &         5325.55 &	 \hbox{Fe \sc ii} &	     E &    5325.56 &	 5325.31   \\
   4840.00 &     \hbox{Fe \sc ii} &          E &    4839.06 &    \nodata &         5333.65 &	 \hbox{Fe \sc ii} &	     E &    5333.81 &	 5331.96   \\
   4848.23 &     \hbox{Cr \sc ii} &          E &    4847.87 &    \nodata &         5336.77 &	 \hbox{Ti \sc ii} &	     E &    5337.19 &	 \nodata   \\
   4856.19 &     \hbox{Cr \sc ii} &          E &    4855.54 &    \nodata &         5376.45 &	 \hbox{Fe \sc ii} &	     E &    5376.32 &	 \nodata   \\
   4862.74 &                    H &          A &    4858.39 &    4859.50 &         5381.02 &	 \hbox{Ti \sc ii} &	     E &    5380.97 &	 \nodata   \\
   4862.74 &                    H &          E &    4861.09 &    4861.27 &         5395.86 &	 \hbox{Fe \sc ii} &	     E &    5395.91 &	 5395.29   \\
   4874.01 &     \hbox{Ti \sc ii} &          E &    4873.45 &    4873.55 &         5402.06 &	 \hbox{Fe \sc ii} &	     E &    5402.05 &	 5401.98   \\	
 \enddata

\end{deluxetable*}

\begin{deluxetable*}{cccccccccc}
\tablecolumns{10}
\tablewidth{0pt}
\tabletypesize{\scriptsize}
\tablenum{2}
\tablecaption{}
\tablehead{   
\colhead{$\lambda_{\mbox{lab}}$} &
\colhead{Species} &
\colhead{Line Type} & 
\colhead{$\lambda_{\mbox{obs}}$} & 
\colhead{$\lambda_{\mbox{obs}}$} &
\colhead{$\lambda_{\mbox{lab}}$} &
\colhead{Species} &
\colhead{Line Type} & 
\colhead{$\lambda_{\mbox{obs}}$} & 
\colhead{$\lambda_{\mbox{obs}}$} \\
\colhead{\ang} &
\colhead{} &
\colhead{} &
\colhead{XX Oph}&
\colhead{AS 325}&
\colhead{\ang}&
\colhead{}&
\colhead{}& 
\colhead{XX Oph}&
\colhead{AS325} \\
\colhead{(1)} &
\colhead{(2)} &
\colhead{(3)} &
\colhead{(4)} & 
\colhead{(5)} & 
\colhead{(6)} &
\colhead{(7)} & 
\colhead{(8)} & 
\colhead{(9)} &
\colhead{(10)}}
\startdata
5414.07 &	  \hbox{Fe \sc ii} &	      E &    5414.00 &    5413.45  &   6746.53 &     \hbox{Fe \sc ii} & 	 E &	6749.99 &    \nodata \\
    5425.26 &	  \hbox{Fe \sc ii} &	      E &    5425.28 &    5425.90  &   6873.84 &     \hbox{Fe \sc ii} & 	 E &	6872.53 &    \nodata \\
    5433.13 &	  \hbox{Fe \sc ii} &	      E &    5433.19 &    \nodata  &   6896.17 &     \hbox{Fe \sc ii} & 	 E &	6895.17 &    \nodata \\
    5465.93 &	  \hbox{Fe \sc ii} &	      E &    5466.59 &    \nodata  &   6922.88 &     \hbox{Fe \sc ii} & 	 E &	6921.49 &    \nodata \\
    5477.24 &	  \hbox{Fe \sc ii} &	      E &    5477.48 &    5477.01  &   7011.23 &     \hbox{Fe \sc ii} & 	 E &	7009.25 &    \nodata \\
    5493.83 &     \hbox{Fe \sc ii} &          E &    5492.85 &    \nodata &    7065.71 &      \hbox{He \sc i} & 	 A &	7064.85 &    \nodata \\
   5502.07 &     \hbox{Cr \sc ii} &          E &    5502.39 &    \nodata &     7065.71 &      \hbox{He \sc i} & 	 E &	7065.61 &    \nodata \\
    5527.61 &     \hbox{Fe \sc ii} &          E &    5527.41 &    5527.99 &     7155.16 &     \hbox{Fe \sc ii} & 	 E &	7154.82 &    7154.63 \\
   5534.85 &     \hbox{Fe \sc ii} &          E &    5535.03 &    5534.66 &     7214.72 &     \hbox{Ti \sc ii} & 	 E &	7214.17 &    7214.17 \\
   5544.76 &     \hbox{Fe \sc ii} &          E &    5544.14 &    \nodata &     7222.39 &     \hbox{Fe \sc ii} & 	 E &	7222.11 &    7221.19 \\
   5586.90 &     \hbox{Fe \sc ii} &          E &    5587.08 &    5586.71 &     7307.97 &     \hbox{Fe \sc ii} & 	 E &	7308.98 &    7309.21 \\
   5627.25 &     \hbox{Fe \sc ii} &          E &    5627.22 &    5628.60 &     7388.18 &     \hbox{Fe \sc ii} & 	 E &	7388.74 &    \nodata \\
   5746.97 &     \hbox{Fe \sc ii} &          E &    5746.79 &    \nodata &     7452.54 &     \hbox{Fe \sc ii} & 	 E &	7450.66 &    7453.12 \\
   5813.68 &     \hbox{Fe \sc ii} &          E &    5813.31 &    \nodata &     7462.38 &     \hbox{Fe \sc ii} & 	 E &	7463.68 &    7464.26 \\
   5823.15 &     \hbox{Fe \sc ii} &          E &    5822.79 &    \nodata &     7479.69 &     \hbox{Fe \sc ii} & 	 E &	7480.93 &    \nodata \\
   5835.45 &     \hbox{Fe \sc ii} &          E &    5835.25 &    5834.91 &     7515.83 &     \hbox{Fe \sc ii} & 	 E &	7516.49 &    7516.74 \\
   5875.60 &      \hbox{He \sc i} &          E &    5875.61 &    \nodata &     7711.72 &     \hbox{Fe \sc ii} & 	 E &	7713.17 &    7714.40 \\
   5889.95 &      \hbox{Na \sc i} &          A &    \nodata &    5888.68 &     7775.39 &       \hbox{O \sc i} & 	 E &	7776.86 &    7778.58 \\
   5895.92 &      \hbox{Na \sc i} &          A &    \nodata &    5894.50 &     7775.40 &       \hbox{O \sc i} & 	 E &	7776.74 &    7778.67 \\
   5902.82 &     \hbox{Fe \sc ii} &          E &    5902.56 &    \nodata &       8316.61 &		      H &	   E &    8316.29 &    8317.66\\
   5991.38 &     \hbox{Fe \sc ii} &          E &    5991.34 &    5990.44 &       8325.77 &		      H &	   E &    8326.44 &    8327.41\\
   6084.11 &     \hbox{Fe \sc ii} &          E &    6084.14 &    6083.42 &       8336.13 &		      H &	   E &    8335.81 &    8337.72\\
   6113.32 &     \hbox{Fe \sc ii} &          E &    6112.85 &    6112.99 &       8347.91 &		      H &	   E &    8347.99 &    8349.06\\
   6129.70 &     \hbox{Fe \sc ii} &          E &    6129.40 &    6129.43 &       8361.36 &		      H &	   E &    8361.20 &    8362.57\\
   6147.73 &     \hbox{Fe \sc ii} &          E &    6148.05 &    6148.03 &       8376.84 &		      H &	   E &    8376.86 &    8378.18\\
   6149.26 &     \hbox{Fe \sc ii} &          E &    6148.60 &    \nodata &       8394.76 &		      H &	   E &    8394.92 &    8396.22\\
   6158.19 &       \hbox{O \sc i} &          E &    6157.72 &    \nodata &       8415.69 &		      H &	   E &    8415.76 &    8416.99\\
   6239.95 &     \hbox{Fe \sc ii} &          E &    6238.61 &    6238.14 &       8440.34 &		      H &	   E &    8440.27 &    8441.31\\
   6247.56 &     \hbox{Fe \sc ii} &          E &    6247.76 &    6246.98 &       8446.80 &	 \hbox{O \sc i} &	   E &    8448.85 &    8449.98\\
   6300.30 &       \hbox{O \sc i} &          E &    6300.29 &    6299.79 &       8469.64 &		      H &	   E &    8469.88 &    8471.41\\
   6317.80 &     \hbox{Fe \sc ii} &          E &    6317.69 &    \nodata &       8498.02 &     \hbox{Ca \sc ii} &	   E &    8500.33 &    8500.87\\
   6331.97 &     \hbox{Fe \sc ii} &          E &    6332.07 &    \nodata &       8504.88 &		      H &	   E &    8504.96 &    8506.40\\
   6369.46 &     \hbox{Fe \sc ii} &          E &    6369.51 &    \nodata &       8542.09 &     \hbox{Ca \sc ii} &	   E &    8544.25 &    8544.76\\
   6383.72 &     \hbox{Fe \sc ii} &          E &    6384.22 &    6382.95 &       8547.79 &		      H &	   E &    8548.12 &    8549.47\\
   6416.92 &     \hbox{Fe \sc ii} &          E &    6416.90 &    6416.74 &       8600.82 &		      H &	   E &    8600.79 &    8602.33\\
   6432.68 &     \hbox{Fe \sc ii} &          E &    6432.55 &    6432.25 &       8662.14 &     \hbox{Ca \sc ii} &	   E &    8664.47 &    8664.96\\
   6456.38 &     \hbox{Fe \sc ii} &          E &    6456.36 &    6455.90 &       8667.46 &		      H &	   E &    8667.82 &    8669.21\\
   6482.31 &     \hbox{Fe \sc ii} &          E &    6482.48 &    \nodata &       8752.94 &		      H &	   E &    8753.31 &    8754.47\\
   6516.08 &     \hbox{Fe \sc ii} &          E &    6516.17 &    \nodata &       8865.28 &		      H &	   E &    8866.04 &    8866.78\\
   6564.71 &                    H &          E &    6563.01 &    6563.99 &       8912.07 &     \hbox{Ca \sc ii} &	   E &    8915.07 &    \nodata\\
   6678.15 &      \hbox{He \sc i} &          A &    6673.13 &    \nodata &       8927.36 &     \hbox{Ca \sc ii} &	   E &    8930.02 &    8929.42\\
   6678.15 &      \hbox{He \sc i} &          E &    6678.23 &    \nodata &       8931.48 &     \hbox{Fe \sc ii} &	   E &    8930.21 &    8931.53\\
\enddata
\tablecomments{ Lab wavelengths listed in columns (1) and (6) were found in the NIST Atomic Spectra Database\tablenotemark{1}. The types are grouped by emission (E) lines and absorption (A) lines.}
\tablenotetext{1}{see http://physics.nist.gov/cgi-bin/AtData/main$\_$asd}

\end{deluxetable*}

\end{document}